# Accurate and fast fiber transfer delay measurement based on phase discrimination and frequency measurement

**J. W. Dong,[1,2] B. Wang,[1,2,*] C. Gao,[1,2] and L. J. Wang[1,2,3,4]**

[1]*Joint Institute for Measurement Science, State Key Laboratory of Precision Measurement Technology and Instruments, Tsinghua University, Beijing 100084, China*
[2]*Department of Precision Instruments, Tsinghua University, Beijing 100084, China*
[3]*Department of Physics, Tsinghua University, Beijing 100084, China*
[4]lwan@tsinghua.edu.cn
[*]bo.wang@tsinghua.edu.cn

**Abstract:** An accurate and fast fiber transfer delay measurement method is demonstrated. As a key technique, a simple ambiguity resolving process based on phase discrimination and frequency measurement is used to overcome the contradiction between measurement accuracy and system complexity. The optimized system achieves a high accuracy of 0.3 ps with a 0.1 ps resolution, and a large dynamic range up to 50 km as well as no dead zone.

**References and links**

1. R.-J. Essiambre, G. Kramer, P. J. Winzer, G. J. Foschini, and B. Goebel, "Capacity limits of optical fiber networks," J. Lightw. Technol. **28**(4), 662-701 (2010).
2. Y. Okawachi, M. S. Bigelow, J. E. Sharping, Z. M. Zhu, A. Schweinsberg, D. J. Gauthier, R. W. Boyd, and A. L. Gaeta, "Tunable all-optical delays via Brillouin slow light in an optical fiber," Phys. Rev. Lett. **94**(15), 153902 (2005).
3. G. Lenz, B. J. Eggleton, C. K. Madsen, and R. E. Slusher, "Optical delay lines based on Optical Filters," IEEE J. Quantum Electron. **37**(4), 525-532 (2001).
4. R. Ramaswamy, K. N. Sivarajan, Optical networks: a practical perspective, (Academic Press, San Francisco, 2002).
5. M. Rost, M. Fujieda, and D. Piester, "Time transfer through optical fibers (TTTOF): progress on calibrated clock comparisons," in Proceedings of 24th European Frequency and Time Forum (2010), paper 6.4.
6. B. Wang, C. Gao, W. L. Chen, J. Miao, X. Zhu, Y. Bai, J. W. Zhang, Y. Y. Feng, T. C. Li, and L. J. Wang, "Precise and continuous time and frequency synchronisation at the 5×10-19 accuracy level," Sci. Rep. **2**, 556 (2012).
7. I. Frigyes and A. J. Seeds, "Optically generated true-time delay in phased-array antennas," IEEE Trans. Microwave Theory Tech. **43**(9), 2378-2386 (1995).
8. W. Ng, A. A. Walston, G. L. Tangonan, J. J. Lee, I. L. Newberg, and N. Bernstein, "The 1st demonstration of an optically steered microwave phased-array antenna using true-time delay," J. Lightwave Technol. **9**(9), 1124–1131 (1991).
9. H.B. Jeon and H. Lee, "Photonic true-time delay for phased-array antenna system using dispersion compensating module and a multiwavelength fiber laser," J. Opt. Soc. Korea **18**(4), 406–413 (2014).
10. K. Predehl, G. Grosche, S. M. F. Raupach, S. Droste, O. Terra, J. Alnis, T. Legero, T. W. Hänsch, T. Udem, R. Holzwarth, and H. Schnatz, "A 920-kilometer optical fiber link for frequency metrology at the 19th decimal place," Science **336**(6080), 441–444 (2012).
11. W. Shillue, "Fiber distribution of local oscillator for Atacama Large Millimeter Array," in Optical Fiber Communication/National Fiber Optic Engineers Conference (IEEE, 2008), pp. 1–3.
12. https://www.ptb.de/emrp/neatft_publications.html.
13. B. Wang, X. Zhu, C. Gao, Y. Bai, J. W. Dong, and L. J. Wang, "Square kilometer array telescope – precision reference frequency synchronisation via 1f-2f dissemination," Sci. Rep. **5**, 13851 (2015).
14. S. D. Personick, "Photon probe: An optical-fiber time-domain reflectometer," Bell Syst. Tech. J. **56**(3), 355–366 (1977).
15. A. Lacaita, P. A. Francese, S. Cova, and G. Ripamonti, "Single-photon optical-time-domain reflectometer at 1.3um with 5cm resolution and high sensivity," Opt. Lett. **18**(13), 1110-1112 (1993).
16. J. Kalisz, "Review of methods for time interval measurements with picosecond resolution," Metrologia **41**(1), 17–32 (2004).

17. B. Qi, A. Tausz, L. Qian, and H. K. Lo, "High-resolution, large dynamic range fiber length measurement based on a frequency-shifted asymmetric Sagnac interferometer," Opt. Lett. **30**(24), 3287–3289 (2005).
18. Y. L. Hu, L. Zhan, Z. X. Zhang, S. Y. Luo, and Y. X. Xia, "High-resolution measurement of fiber length by using a mode-locked fiber laser configuration," Opt. Lett. **32**(12), 1605–1607 (2007).
19. K. Yun, J. Li, G. Zhang, L. Chen, W. Yang, and Z. Zhang, "Simple and highly accurate technique for time delay measurement in optical fibers by free-running laser configuration," Opt. Lett. **33**(15), 1732–1734 (2008).
20. S. Y. Set, M. Jablonski, K. Hsu, C. S. Goh, and K. Kikuchi, "Rapid amplitude and group-delay measurement system based on intra-cavity-modulated swept-lasers," IEEE Trans. Instrum. Meas. **53**(1), 192–196 (2004).
21. K. Yoon, J. Song, and H. D. Kim, "Fiber length measurement technique employing self-seeding laser oscillation of Fabry-Perot laser diode," Jpn. J. Appl. Phys. **46**(1), 415 (2007).
22. J. W. Dong, B. Wang, C. Gao, Y. C. Guo, and L. J. Wang, "Highly accurate fiber transfer delay measurement with large dynamic range," Opt. Express **24**(2), 1368–1375 (2016).

## 1. Introduction

The fiber transfer delay (FTD), as an essential characteristic for fiber transmission, becomes a significant parameter in many optical fiber applications, such as optical communication [1–3], fiber timing distribution system [4–6], phased array antenna [7–9] and some other large-scale scientific or engineering facilities [10–13]. Up to now, many techniques for FTD measurement in time domain or frequency domain have been developed and successfully employed [14-21]. However, most of them either have drawbacks of low accuracy and existing dead zones, or require complex and time-consuming process. Pulse methods using optical time-domain reflectometer (OTDR) or time interval counter (TIC) are conventional methods still in use due to its efficiency and flexibility [14–16]. In these methods, the FTD is determined by measuring the time-of-flight, or group delay of narrow pulses with a resolution of few tens of picoseconds. However, susceptible to the pulse width and the band-limited receiver, it may no longer meet the growing accuracy requirements nowadays. New approaches also have been evolving in which FTD is measured using rather sophisticated and expensive equipment, such as mode-locked fiber laser or self-seeding laser oscillator [17–21]. They achieve a fairly high accuracy, yet inevitably induce more time-consumption for using complex ambiguity resolving process.

We have recently developed an FTD measurement method [22]. By transferring a laser light modulated by a microwave signal and locking the frequency of the signal onto the FTD, the FTD measurement is converted into frequency measurement. In addition, the conventional pulse method is used for resolving ambiguity. The FTD measurement accuracy of 1 ps and resolution of 0.2 ps are obtained, together with a dynamic range over 50 km as well as no dead zone. For ambiguity resolving, the additional coarse FTD measurement need to be accurate enough. While due to the pulse width limitation and inevitable pulse broadening, the accuracy of the conventional pulse method deteriorates when the length of the fiber under test (FUT) increases. Moreover, the pulse signal transmission makes a complex equipment and a time-consuming measurement procedure.

In this paper, we demonstrate an important optimization of the method. Concerning the system complexity and time consuming owing to the coarse FTD measurement, a new technique for ambiguity resolving process is proposed. Replacing pulse signal transmission and the corresponding TIC instruments, the ambiguity is resolved via fast phase detection and frequency measurement operations. This new technique greatly simplifies the measurement procedure and requires only low-cost and much simpler equipment. The measurement results indicate that the improved FTD measurement system achieves a very high accuracy of 0.3 ps with 0.1 ps resolution, and a large dynamic range of up to 50 km as well as no dead zone. The accuracy comparison measurements have also been carried out between the optimized method and the original pulse method.

## 2. Methods

Figure 1 shows the schematic diagram of the optimized FTD measurement system which is composed of FTD measurement loop and system delay control (SDC) loop. In FTD measurement loop (Fig. 1(a)), via an optical circulator, the 1550 nm laser light modulated by a microwave signal is transferred in the fiber under test (FUT). The microwave signal approximately 1 GHz is generated by a programmable frequency synthesizer phase-locked to a 100 MHz voltage control oscillator (VCO). A Faraday mirror is connected to one end of the FUT to reflect the light for detection. After frequency mixing and filtering operations on the signals before and after transmission, an error signal $V_e \propto \cos(\phi_p)$ is obtained, where $\phi_p$ denotes the transmission phase delay. Thus $\phi_p$ can be expressed as

$$\phi_p = 2\pi f \cdot t. \tag{1}$$

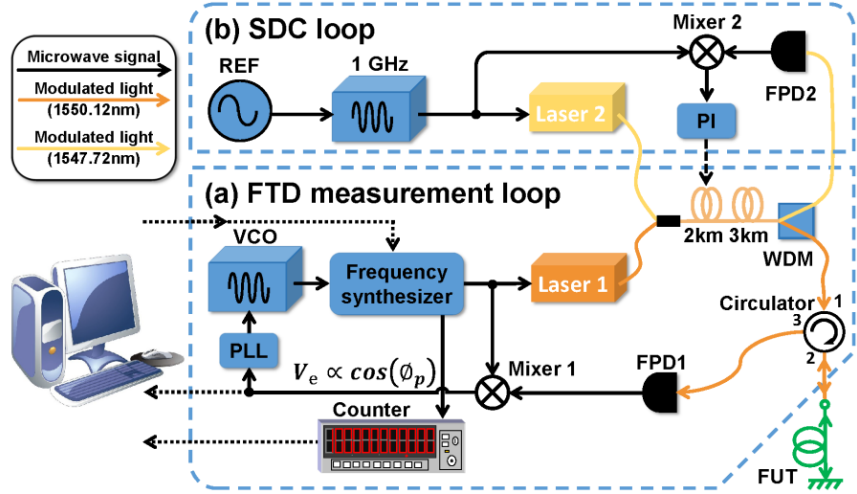

Fig. 1. Schematic of the optimized FTD measurement system: (a) FTD measurement loop. (b) SDC loop. VCO, 100 MHz voltage control oscillator; FPD, fast photo detector with a typical 3dB bandwidth of 16 GHz; FUT, fiber under test; PLL, phase locked loop; WDM, 1547/1550 nm wavelength division multiplexer; REF, reference signal; PI, proportional integral controller.

Here, $f$ denotes the frequency of the microwave signal which can be measured by a frequency counter; $t$ denotes transfer delay, which contains the double-passed FUT delay and the system delay (including propagation delays of electronic circuits, cables, and fiber pigtails in the FTD measurement loop). A phase locked loop (PLL) is used to control the frequency of VCO, yielding $V_e \propto \cos(\phi_p) = 0$. Then, the following equation is valid:

$$\phi_p = (N + \frac{1}{2})\pi. \tag{2}$$

Here, $N$ is an integer. Based on Eq. (1) and (2), the derived equation is

$$t = \frac{2N+1}{4f}. \tag{3}$$

When the PLL is locked, $N$ is fixed as a constant, making frequency of the microwave signal $f$ locked onto the transfer delay $t$. Here, $N$ can be considered as an ambiguity of the measurement. In the previous method [22], a coarse FTD ($t_{coarse}$) measurement using conventional pulse method is employed to resolve this ambiguity. For an operational frequency at 1 GHz, the coarse FTD measurement uncertainty should be below 250 ps. Here, we used a new method for coarse FTD measurement. For the same FTD, the locked frequency $f$ is a separated variable corresponding to different values of $N$. Through changing the multiple of frequency synthesizer and performing two frequency locking operations, we get two different frequencies $f_1$ and $f_2$, accompany with their ambiguities $N_1$ and $N_2$ respectively. And the following relationship is valid:

$$t = \frac{2N_1+1}{4f_1} = \frac{2N_2+1}{4f_2}. \tag{4}$$

The coarse FTD can be calculated by

$$t_{coarse} = \frac{N_2-N_1}{2(f_2-f_1)}. \tag{5}$$

Here, $f_1$ and $f_2$ can be obtained by frequency measurement. In order to get $(N_2-N_1)$, the frequency synthesizer is programmed to make the frequency of the microwave signal scanning from $f_1$ to $f_2$ in bursts of $\delta f$ steps. Meanwhile, the frequency scanning caused phase delay shift is monitored by recording the error signal $V_e \propto \cos(\phi_p)$. In this way, $(N_2-N_1)$ can be obtained from the zero-crossing detection of $V_e$.

Additionally, according to Eq. (1), the frequency increment ($\delta f$) induced phase delay shift ($\delta\phi_p$) can be expressed as

$$\delta\phi_p = 2\pi \cdot \delta f \cdot t. \tag{6}$$

To acquire all zero crossing points, each frequency step induced phase delay shift should not exceed $\pi/2$. Under the FUT length below 50 km, the FTD will not exceed 0.5 ms, resulting in a maximum frequency scanning step of 500 Hz. To eliminate the measurement dead zone, a 5 km long fiber is inserted into the system [22].

With the determined $(N_2-N_1)$, $f_1$ and $f_2$, according to Eq. (5), $t_{coarse}$ can be calculated. The uncertainty of $t_{coarse}$ can be analyzed as

$$\Delta t_{coarse} = \frac{(N_2-N_1)}{2(f_2-f_1)^2} \cdot \Delta(f_2-f_1) = t_{coarse} \cdot \frac{\Delta(f_2-f_1)}{f_2-f_1}. \tag{7}$$

Here, $\Delta(f_2-f_1)$ is the uncertainty of the frequency difference. It is mainly caused by the measurement uncertainty of the two frequencies and the FTD variation during two frequency locking operations. For the former, the measurement uncertainty of a commercial meter is better than $10^{-10}$, corresponding to 0.1 Hz frequency uncertainty for a 1 GHz carrier frequency. For the latter, due to the fast frequency locking operation, the time interval could be less than 2 s, making the FTD variation-induced frequency shift not to exceed 3 Hz (corresponding to 10℃/hour temperature variation). According to [22], under the operation frequency of 1 GHz, $\Delta t_{coarse} < 250$ ps is required for accurate ambiguity resolving. Consequently, for FTD below 0.5 ms (50 km FUT), the frequency scanning range $(f_2-f_1)$ should be more than 6 MHz. Since the frequency synthesizer has a scanning range up to 10 MHz, the coarse FTD measurement is accurate enough. Once $t_{coarse}$ is obtained with an uncertainty of below 250 ps, both $N_1$ and $N_2$ can be precisely determined as following,

$$N_1 = 2f_1 t_{coarse} - \frac{1}{2}, \tag{8}$$

$$N_2 = 2f_2 t_{coarse} - \frac{1}{2}. \tag{9}$$

In this way, the ambiguity is resolved. When the frequency is locked, $t$ can be obtained through frequency measurement according to Eq. (3).

In addition, when no FUT is connected, the laser light is reflected by the end surface of the circulator (port 2, FC/PC connector). System delay ($t_0$) is calibrated via the same method and the one-way FUT delay ($t_F$) can be obtained as

$$t_F = \frac{1}{2}(t-t_0) = \frac{1}{2}\left(\frac{2N_2+1}{4f_2} - t_0\right). \tag{10}$$

Meanwhile, a system delay control (SDC) loop, which is the same as the previous method [22], is added to compensate the system delay fluctuation caused by temperature variation (shown in Fig. 1(b)). Once the system delay is calibrated and $N_1$ ($N_2$) is determined and fixed as a constant, the long-term, real-time FTD measurement of the FUT $t_F$ can be implemented by continuous measurement of the frequency $f_1$ ($f_2$).

## 3. Results and discussion

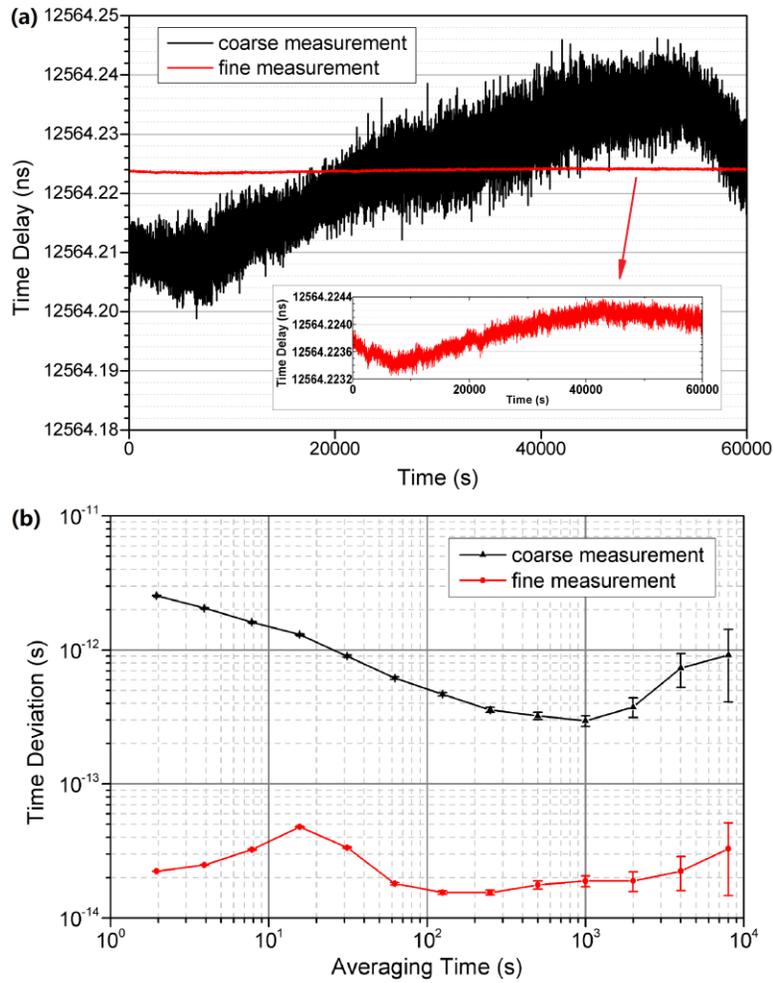

Fig. 2. Measurement results of the compensated system delay using coarse FTD method and fine FTD method. (a) The compensated system delay. The black line is the result of coarse FTD measurement, showing a fluctuation of ±20 ps. The red line is the result of fine FTD measurement, showing a fluctuation of ±0.6 ps. During the coarse measurement,

the frequency scans from 995.028 MHz to 1004.978 MHz in bursts of 1000 Hz. (b) The time deviation derived from the measured compensated system delay.

To verify the accuracy of the proposed coarse measurement method, alternate measurements of the compensated system delay without connecting FUT using coarse FTD method and fine FTD method have been performed. Figure 2 shows the measured system delay and the converted system stability via time deviation (TDEV). The black line is the coarse measurement results, and the red line represents the fine measurement results. In Fig. 2(a), we observed the fluctuation of ±20 ps in coarse measurement and ±0.6 ps in the fine measurement mode. The fluctuations can also be clearly observed in the corresponding TDEV plot, shown in Fig. 2(b). The values of fine measurement are always below 50 fs at different averaging times, whereas in the coarse measurement, the TDEV reaches 2.5 ps at minimum measurement time (about 2 s). It also indicates that the accuracy of the proposed coarse measurement is sufficient for resolving the ambiguity.

To further evaluate the accuracy of the optimized FTD measurement method, we used it to measure the FTD of a 2 m long fiber which can be considered as a constant. Comparison measurements have also been demonstrated using the conventional pulsed method. We repeat tests at different times of day and the system delay calibration has been done before each test. The results are shown in Fig. 3. It can be seen that the mean values of all measured FTDs using both methods agree very well. In the measurement using commercial TIC, the statistical error reaches approximately 13 ps and the mean value fluctuation is around ±5 ps. In contrast, the statistical error of the optimized method is only 0.1 ps, which reflects the resolution of the measurement system. And the mean value fluctuation decreases to ±0.3 ps, reflecting the accuracy of the measurement system.

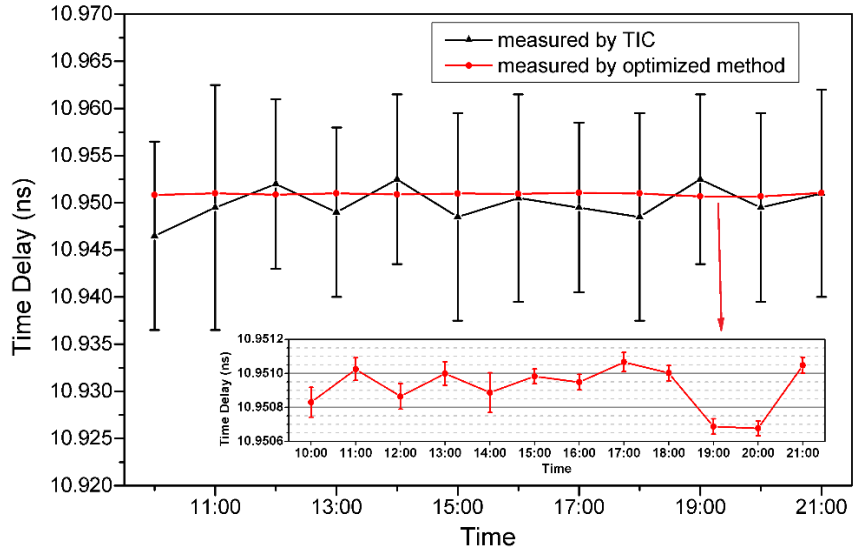

Fig. 3. FTD measurement results of a 2 m long fiber. The averaging period is 20 s. The error bar is the standard deviation of each measurement. The black line is the FTD measured by TIC, showing a mean value fluctuation of ±5 ps with a statistical error of 13 ps. The red line is the FTD measured by optimized method. The mean value fluctuation is reduced to ±0.3 ps with a statistical error of 0.1 ps.

To demonstrate the measurement range of the optimized FTD measurement method, we also measure the FTD of a 50 km fiber spool. Comparison tests using two methods have also been carried out at different times of day. The results are shown in Fig. 4. It can be seen that, the mean values using two methods are well overlapped. The long-term variation of 1100 ps is caused by temperature variation during the measurement time. When using a commercial TIC, the statistical error increases to 100 ps because of the pulse broadening and nonlinear optical distortion. On the other hand, by using the optimized method, the statistical error is below 0.4 ps. It indicates that the optimized method also has a large dynamic range of at least up to 50 km, while at the mean time preserving an extremely high resolution.

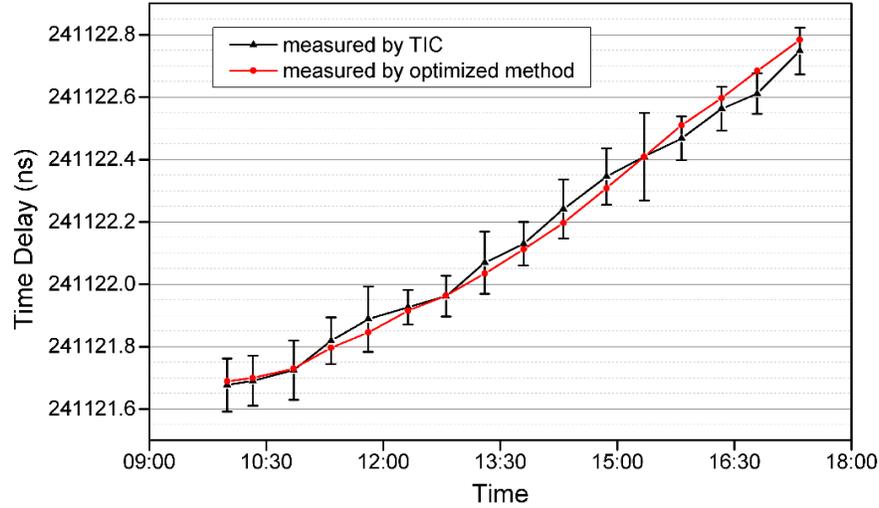

Fig. 4. FTD measurement results of a 50 km long fiber. The averaging period is 10 s. The error bar is the standard deviation of each measurement. The long-term fluctuation is around 1100 ps. The black line is the FTD measured by TIC, showing a statistical error below 100 ps. The red line is the FTD measured by optimized method, the statistical error of which is around 0.4 ps.

### 4. Conclusion

We have optimized the previous FTD measurement system using an accurate and fast FTD measurement method as the ambiguity resolving process. Using the proposed method, the FTD can be coarsely determined via a phase shift and frequency locking technique within only a few seconds. This optimization greatly simplifies the measurement procedure and requires only low-cost and much simpler equipment. As a result, the improved system can achieves a very high accuracy of 0.3 ps with a 0.1 ps resolution, and a large dynamic range of up to 50 km. A good agreement in measurement results with the conventional pulsed method is obtained as well. Using this new method, FTD measurement no longer need to rely on the conventional pulsed method and it can become much more convenient and faster.